\documentclass[aps,pra,twocolumn,10pt]{revtex4-2}
\usepackage{amsmath}
\usepackage{amssymb}
\usepackage{amsthm}
\usepackage{mathdots}
\usepackage{bm}
\usepackage{bbm}
\usepackage{graphicx}
\usepackage{epstopdf}
\usepackage{tikz}
\usepackage{amscd}
\usepackage{booktabs}
\usepackage{tabularx}
\usepackage{dcolumn}
\usepackage{multirow}
\usepackage{adjustbox}
\usepackage{enumerate}
\usepackage{changepage}
\usepackage{xcolor}
\usepackage{mathrsfs}
\usepackage[
colorlinks,
linkcolor = blue,
citecolor = blue,
urlcolor = blue
]{hyperref}
\newtheorem{theorem}{Theorem}
\newtheorem{proposition}{Proposition}
\newtheorem{corollary}{Corollary}
\newtheorem{lemma}{Lemma}

\newtheorem{definition}{Definition}

\usepackage{graphicx}% Include figure files
\usepackage{bm}% bold math
\begin{document}

\preprint{APS/123-QED}

\title{Quantum Causal Compatibility from Spatiotemporal Correlations}% Force line breaks with \\

\author{Ya-Xi Ban$^{1,2}$} 
\author{Zhu-Jun Zheng$^{1，2}$} \email{zhengzj@scut.edu.cn}
\author{Shao-Ming Fei$^{3}$}
\author{Chang-Yue Zhang$^{4}$}
\email{cyzhang@jnu.edu.cn}
\author{Mao-Sheng Li$^{1}$} 
\affiliation{ 
\small $^{1}$ School of Mathematics, South China University of Technology, Guangzhou 510640, People's Republic of China\\
\small $^{2}$ Laboratory of Quantum Science and Engineering, South China University of Technology, Guangzhou 510641, People's Republic of China\\
\small $^{3}$ School of Mathematical Sciences, Capital Normal University, Beijing 100048, People's Republic of China\\
\small $^{4}$ School of Mathematics, Jinan University, Guangzhou 510632, People's Republic of China
}

\begin{abstract}
Which quantum causal structures can account for observed spatiotemporal correlations without access to an intervention-complete description of the underlying process? In particular, can such correlations distinguish a temporal chain, a common cause, and a common effect? Here, we address these questions using three-qubit pseudo-density operators constructed from experimentally accessible Pauli correlations. We derive necessary and sufficient conditions for pseudo-Markov compatibility with the chain, fork, and collider structures by relaxing complete positivity while preserving the corresponding quantum Markov constraints. Despite their distinct causal mechanisms, all three pseudo-Markov compatibility problems reduce to linear feasibility tests, without requiring full-rank marginal states. We further establish a unified semidefinite-programming approach to physical realizability and quantify the minimum violation of complete positivity through Choi negativity. Using a two-parameter family of pseudo-density operators, we demonstrate that different causal structures admit overlapping yet non-nested sets of correlations. Our results show that quantum causal explanations need not be unique, but experimentally accessible correlations can nevertheless distinguish and exclude candidate causal structures.
\end{abstract}

\maketitle

\section{Introduction}
Causal inference, understood as the task of identifying cause-and-effect relations from observed correlations, is a fundamental problem across the sciences. In the classical setting, Bayesian networks provide a systematic framework for describing causal dependencies among multiple variables, testing candidate causal structures, and predicting the consequences of interventions~\cite{Pearl2009, Spirtes2000}. Recent advances in the preparation, control, and measurement of quantum systems have stimulated growing interest in extending causal inference to the quantum domain~\cite{Ried2015, MacLean2017, Carvacho2017, Chaves2018, Agresti2022}. Such an extension requires a careful reconsideration of central classical notions, including observation, intervention, conditional independence, and the causal Markov condition, which has motivated the development of quantum causal models, quantum conditional states, and generalized Bayesian networks~\cite{Leifer2013, Henson2014, PienaarBrukner2015, Costa2016, Allen2017, Pienaar2020}. At the same time, quantum theory permits causal phenomena without direct classical counterparts, including coherent combinations of common-cause and direct-cause mechanisms~\cite{MacLean2017, Feix2017, Gachechiladze2020}, indefinite causal order~\cite{Oreshkov2012, Brukner2014, Chiribella2013, OreshkovGiarmatzi2016}, and multi-time correlations mediated by quantum memory~\cite{Pollock2018, PollockPRA2018, Taranto2019, White2020, MilzModi2021}. The identification, certification and exclusion of candidate causal structures from experimentally accessible quantum data have therefore become central objectives of quantum causal inference~\cite{Ried2015, Costa2016, Giarmatzi2018, Javidian2022, FriendKissinger2024, Liu2025LightTouch}.

Quantum causal relations are often studied at the level of quantum processes. Quantum combs~\cite{Chiribella2009} and process tensors~\cite{PollockPRA2018,MilzModi2021} characterize multitime dynamics through their responses to quantum operations, providing tools for studying memory effects~\cite{Pollock2018,Taranto2019,White2020} and causal structures~\cite{Bai2022}. Process matrices~\cite{Oreshkov2012} accommodate scenarios without a predefined causal order, while quantum causal models~\cite{Costa2016,Allen2017} impose structural constraints associated with candidate causal graphs. These frameworks also support quantum causal discovery~\cite{Giarmatzi2018}. In parallel, state-like descriptions of temporal and spatiotemporal correlations have been developed, including multiple-time states~\cite{Aharonov2009} and quantum states over time~\cite{Horsman2017,FullwoodParzygnat2022,LieNg2024,Fullwood2025,LieFullwood2025}. These approaches seek to extend state-based descriptions to events at different times, providing alternative perspectives on temporal quantum correlations. In particular, the pseudo-density operator (PDO) formalism encodes Pauli correlations in a single Hermitian  unit-trace operator that need not be positive semidefinite~\cite{Fitzsimons2015,Horsman2017}. Since quantum measurements can disturb subsequent dynamics~\cite{Leggett1985,Ku2018}, and experimental data may be restricted to correlations obtained under a fixed measurement scheme, PDOs provide a useful starting point for investigating causal compatibility directly from the available Pauli data, without requiring an intervention-complete process description.

The PDO framework has been used to study temporal quantum correlations~\cite{Fitzsimons2015,Horsman2017,Ku2018,LiuTemporal2025}, their connections to quantum dynamics~\cite{Utagi2021,Fullwood2025}, and temporal directionality~\cite{Liu2024Arrow}. Recent developments include the causal classification of bipartite spatiotemporal correlations~\cite{Song2024,Liu2025LightTouch} and related investigations of direct causal explanations of bipartite quantum states~\cite{SongParzygnat2025}. At the multipartite level, related studies have considered spacetime marginal constraints~\cite{Jia2023SpacetimeMarginal}, Markovian extensions of quantum states over time~\cite{LieFullwood2025}, and causal compatibility in quantum networks~\cite{Smith2026Inflation}. For multipartite PDOs, however, a central challenge is to determine whether the encoded Pauli correlations admit a globally consistent representation satisfying the Markov constraints of a prescribed causal graph. In particular, compatibility of individual two-event marginals does not generally guarantee consistency with the remaining pairwise and genuine multipartite correlations.

\begin{figure}[t]
\centering
\includegraphics[width=1\columnwidth]
{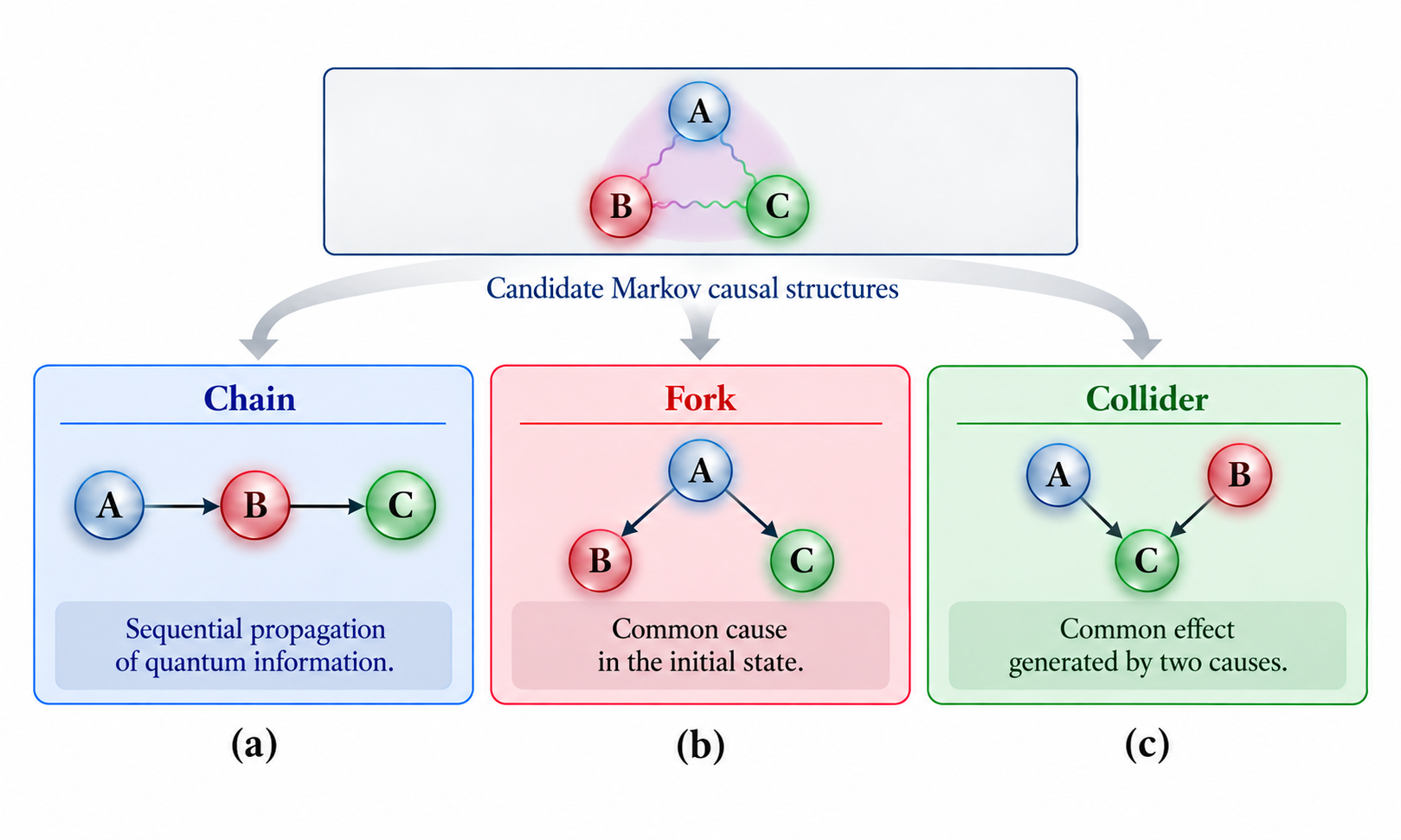}
\caption{
Candidate causal structures for three quantum events $A$, $B$, and $C$: (a) a chain, describing sequential causal propagation; (b) a fork, representing a common cause; and (c) a collider, representing a common effect. Undirected lines indicate observed correlations without specifying causal directions, while directed arrows represent candidate causal relations. One representative assignment of the event labels is shown for each topology.
}

\label{fig:causal_structures}
\end{figure}

In this work, we address this global compatibility problem for three-qubit PDOs by considering the chain, fork, and collider causal structures illustrated in Fig.~\ref{fig:causal_structures}, including their different node assignments~\cite{Pearl2009,Spirtes2000,Lauand2023}. By separating the algebraic constraints of quantum Markov structures from the requirement of complete positivity, we introduce pseudo-Markov compatibility in terms of Hermiticity-preserving trace-preserving (HPTP) mechanisms. Using the observed Pauli data, we derive necessary and sufficient conditions for pseudo-Markov compatibility with each structure and show that the resulting compatibility problems reduce to linear feasibility. We further develop a unified semidefinite-programming framework for testing physical realizability and introduce a graph-dependent nonphysicality measure based on the minimum total Choi negativity of compatible mechanisms. Finally, a two-parameter PDO family demonstrates that the physical compatibility regions of different causal structures can overlap, while a representative PDO illustrates how the algebraic compatibility criteria and the graph-dependent nonphysicality measure can be used to distinguish different causal structures from the same observed correlations.

The paper is organized as follows. In Sec.~II, we introduce the PDO formalism and formulate the causal compatibility framework. In Sec.~III, we derive the pseudo-Markov compatibility conditions for chain structures and discuss the reconstruction of compatible pseudo-channels. Sections~IV and V extend the analysis to fork and collider structures, respectively. In Sec.~VI, we investigate physical realizability through complete positivity and introduce a graph-dependent nonphysicality measure based on Choi negativity. In Sec.~VII, we illustrate the compatibility regions of different causal structures using a two-parameter family of three-qubit PDOs and examine a representative example. Finally, Sec.~VIII summarizes our results and discusses possible extensions.

\section{PSEUDO-DENSITY OPERATORS AND CAUSAL COMPATIBILITY}
\label{sec:framework}
Consider three events $A$, $B$ and $C$, each associated with a qubit. In each experimental round, Pauli settings $a,b,c\in\{0,1,2,3\}$ are chosen at the three events. Here $\sigma_1,\sigma_2,\sigma_3$ denote the standard Pauli operators, and $\sigma_0=I$. A nonzero setting corresponds to a projective Pauli measurement with outcomes $\pm1$ and the associated L\"uders state update, whereas setting $0$ denotes no measurement, with its outcome factor assigned the value $+1$. Let $P(x,y,z|a,b,c)$ denote the observed joint distribution. The Pauli correlators are
\begin{align*}
    r_{abc}:=\langle\sigma_a,\sigma_b,\sigma_c\rangle
=\sum_{x,y,z=\pm1}xyz\,P(x,y,z|a,b,c).
\end{align*}
These correlators determine the three-qubit PDO~\cite{Fitzsimons2015},
\begin{equation}
R_{ABC}=\frac18\sum_{a,b,c=0}^{3}
r_{abc}\,\sigma_a\otimes\sigma_b\otimes\sigma_c.
\label{R_ABC}
\end{equation}
By construction, $R_{ABC}$ is Hermitian and has unit trace since $r_{000}=1$, but it need not be positive semidefinite.

We represent $R_{ABC}$ by its Pauli data $(\boldsymbol u,\boldsymbol v,\boldsymbol w,M,N,K,G)$. For $a,b,c\in\{1,2,3\}$, the one-body Bloch vectors are $u_a=r_{a00}$, $v_b=r_{0b0}$, and $w_c=r_{00c}$; the two-body correlation matrices are $M_{ab}=r_{ab0}$, $N_{bc}=r_{0bc}$, and $K_{ac}=r_{a0c}$; and the three-body correlation tensor is $G_{abc}=r_{abc}$. Thus, $\boldsymbol u,\boldsymbol v,\boldsymbol w\in\mathbb R^3$, $M,N,K\in\mathbb R^{3\times3}$, and $G\in\mathbb R^{3\times3\times3}$. Together with $r_{000}=1$, these quantities completely specify $R_{ABC}$.

We now consider quantum causal models (QCMs) associated with directed acyclic graphs (DAGs)~\cite{Allen2017}. Let $\mathsf G=(V,E)$ be a finite DAG with vertex set $V=\{A_1,\ldots,A_n\}$ and directed edge set $E$, and let $\operatorname{Pa}_{\mathsf G}(A_i)$ denote the parent set of node $A_i$. Each node is assigned input and output Hilbert spaces, corresponding to the system before and after a local intervention. A QCM specifies, for each node $A_i$, a conditional Choi operator $J_{A_i|\operatorname{Pa}_{\mathsf G}(A_i)}$ representing a completely positive trace-preserving (CPTP) map from the joint output systems of its parents to the input system of $A_i$. In the unnormalized Choi convention, these operators satisfy $J_{A_i|\operatorname{Pa}_{\mathsf G}(A_i)}\succeq 0$ and $\operatorname{Tr}_{A_i^{\mathrm{in}}}J_{A_i|\operatorname{Pa}_{\mathsf G}(A_i)}=I_{\operatorname{Pa}_{\mathsf G}(A_i)^{\mathrm{out}}}$. For a root node, $J_{A_i|\varnothing}=\rho_{A_i}$ is its preparation state. The conditional Choi operators are required to commute pairwise when embedded in the common process space. The corresponding process operator is
\begin{equation}
\Sigma_{\mathsf G}
=\prod_{i=1}^{n}
J_{A_i|\operatorname{Pa}_{\mathsf G}(A_i)}.
\label{eq:general-Markov-factorization}
\end{equation}
Throughout, identity operators on spectator systems are implicit. A valid quantum process operator is said to be Markov for $\mathsf G$ if and only if it admits such a factorization into pairwise commuting local CPTP channel operators ~\cite{Allen2017,Barrett2021}.

Unlike a process operator, which acts on separate input and output spaces for each event, a PDO retains a single operator factor per observed event. The relation between process matrices and PDOs has been investigated in Ref.~\cite{Liu2024}. Here, we consider projective measurements of Pauli observables with L\"uders state updates, with the identity setting corresponding to no measurement. This measurement protocol defines a map $\Phi_{\mathrm{PL}}$ from a process operator to the corresponding PDO. For a process operator $\Sigma$ describing $n$ qubit events, the map is given by
\begin{equation}
\Phi_{\mathrm{PL}}(\Sigma)=\frac{1}{2^n}\sum_{\alpha_1,\ldots,\alpha_n=0}^{3}r_{\alpha_1\cdots\alpha_n}[\Sigma]\,
\sigma_{\alpha_1}\otimes\cdots\otimes\sigma_{\alpha_n},
\label{eq:process-to-PDO-map}
\end{equation}
where $r_{\alpha_1\cdots\alpha_n}[\Sigma]$ denotes the corresponding Pauli correlation function, evaluated using the generalized Born rule for the prescribed local measurement instruments. This connection allows us to formulate causal compatibility at the level of observed PDOs by asking whether their Pauli correlations can be reproduced by a quantum Markov process associated with a given DAG.

\begin{definition}[Markov compatibility]
\label{def:markov_compatibility}
A three-qubit PDO $R_{ABC}$ is said to be Markov compatible with a DAG $\mathsf G$ on $\{A,B,C\}$ if there exist quantum states for the root nodes and CPTP maps for the non-root nodes such that the resulting quantum process is Markov for $\mathsf G$ and reproduces all Pauli correlations encoded in $R_{ABC}$ under the prescribed Pauli measurement scheme with L\"uders state updates.
\end{definition}

Equivalently, there exists a valid process operator $\Sigma_{\mathsf G}$ of the form \eqref{eq:general-Markov-factorization} such that $R_{ABC}=\Phi_{\mathrm{PL}}(\Sigma_{\mathsf G})$. We further introduce pseudo-Markov compatibility by relaxing complete positivity of the non-root local mechanisms while retaining physical root preparations and the graph-dependent factorization and commutativity constraints. The CPTP maps are thus replaced by HPTP pseudo-channels. A PDO is pseudo-Markov compatible with $\mathsf G$ if these algebraically admissible mechanisms reproduce its Pauli correlations under the formal extension of the prescribed Pauli measurement rules. The resulting pseudo-realization need not define a valid physical quantum process.

In the following, we specialize to three qubit events and investigate the chain, fork, and collider structures, including their different node assignments. Our goal is to derive the necessary and sufficient conditions for pseudo-Markov compatibility directly from the observed Pauli data
$(\boldsymbol u,\boldsymbol v,\boldsymbol w,M,N,K,G)$.

\section{Chain Causal Compatibility}
We first consider the chain structure $\mathsf G_{\mathrm{ch}}:A\to B\to C$, in which the causal influence of $A$ on $C$ is entirely mediated through the intermediate event $B$. Although $A$ and $C$ are not directly connected, they may still exhibit correlations arising from this indirect causal influence. The chain structure requires these correlations, together with the three-body correlations, to be consistent with the same pair of successive causal mechanisms. To characterize this requirement, consider a qubit initially prepared in state $\rho_A$ and evolving through two successive CPTP channels $\mathcal E_{B|A}$ and $\mathcal E_{C|B}$. Under the Pauli measurement scheme introduced in Sec.~\ref{sec:framework}, the joint probability distribution is

\begin{equation}
\begin{aligned}
P(x,y,z|a,b,c)={}&\operatorname{Tr}\Big[\Pi_{z|c}\mathcal E_{C|B}\Big(\Pi_{y|b}\\
&\quad\mathcal E_{B|A}\big(\Pi_{x|a}\rho_A\Pi_{x|a}\big)\Pi_{y|b}\Big)\Big].
\end{aligned}
\label{pro}
\end{equation}
where $\Pi_{x|a}=(I+x\sigma_a)/2$ for $a=1,2,3$, with $\Pi_{+|0}=I$ and $\Pi_{-|0}=0$, and analogous conventions at $B$ and $C$. The resulting three-event PDO admits the following representation.

\begin{proposition}
\label{prop:R}
For a qubit state $\rho_A$ and two successive CPTP channels $\mathcal E_{B|A}$ and $\mathcal E_{C|B}$ in order $A\to B\to C$, the corresponding three-qubit PDO is
\begin{equation}
\begin{split}
R_{ABC}=\frac1{16}\sum_{a,b=0}^{3}\sigma_a\otimes\sigma_b\otimes\mathcal E_{C|B}\Big(\big\{\mathcal E_{B|A}(\{\rho_A,\sigma_a\}),\sigma_b\big\}\Big).
\end{split}
\label{eq:physical-chain-pdo}
\end{equation}
\end{proposition}

The proof is given in Appendix~\ref{prop1:proof}. Equation~\eqref{eq:physical-chain-pdo} provides a forward representation of the observed PDO in terms of the initial state and two successive causal mechanisms. The causal compatibility problem concerns the converse: given $R_{ABC}$, can one find mechanisms that reproduce all its Pauli correlations while respecting the proposed chain structure? As discussed in Sec.~\ref{sec:framework}, we first address this question at the level of pseudo-Markov compatibility, where the CPTP channels are replaced by HPTP pseudo-channels. A natural starting point is the two adjacent marginals $R_{AB}$ and $R_{BC}$, since any sequential realization must reproduce both. For operational two-qubit PDOs, the existence of compatible pseudo-channels is guaranteed by the following result.

\begin{lemma}~\cite{Song2024}
\label{2PDO-compatible}
Every operational two-qubit PDO admits at least one compatible forward and reverse HPTP pseudo-channel.
\end{lemma}

Nevertheless, the separate compatibility of $R_{AB}$ and $R_{BC}$ does not ensure that their corresponding mechanisms can be combined to reproduce the full three-event PDO. The nonadjacent $AC$ correlations and genuine three-body correlations impose additional requirements that are invisible to the adjacent marginals alone. To identify these requirements, we first determine how the observed two-event data constrain the underlying pseudo-channel. Using the real Bloch-affine representation of HPTP qubit maps~\cite{Settimo2022,ChoiLi2023}, $\boldsymbol r\mapsto T\boldsymbol r+\boldsymbol t$, with $T\in\mathbb R^{3\times3}$ and $\boldsymbol t\in\mathbb R^3$, we obtain the following reconstruction result.

\begin{lemma}
\label{lem:pseudo-channel-reconstruction}
Let $R_{AB}$ be a two-qubit PDO with Pauli data $(\boldsymbol u,\boldsymbol v,M)$. Any compatible HPTP pseudo-channel $\mathcal E_{B|A}$ with Bloch-affine parameters $(T,\boldsymbol t)$ satisfies
\begin{equation}
M=T^{\mathsf T}+\boldsymbol u\boldsymbol t^{\mathsf T},~~\boldsymbol v=T\boldsymbol u+\boldsymbol t.
\label{eq:two-time-compatibility}
\end{equation}
Conversely, any real parameters satisfying these relations define a compatible HPTP pseudo-channel.

If $\|\boldsymbol u\|<1$, the compatible pseudo-channel is unique, with
\begin{equation}
\boldsymbol t=\frac{\boldsymbol v-M^{\mathsf T}\boldsymbol u}{1-\|\boldsymbol u\|^2},~~T=M^{\mathsf T}-\boldsymbol t\boldsymbol u^{\mathsf T}.
\label{eq:two-time-full-rank}
\end{equation}
If $\|\boldsymbol u\|=1$, compatibility requires $\boldsymbol v=M^{\mathsf T}\boldsymbol u$. When this condition holds, all compatible pseudo-channels are parameterized by
\begin{equation}
\boldsymbol t\in\mathbb R^3,~~T(\boldsymbol t)=M^{\mathsf T}-\boldsymbol t\boldsymbol u^{\mathsf T}.
\label{eq:two-time-rank-deficient-family}
\end{equation}
\end{lemma}

The proof is given in Appendix~\ref{lemma2:proof}. The lemma shows that a full-rank input marginal uniquely determines the compatible pseudo-channel, whereas a rank-deficient marginal may leave an affine family of admissible maps. In the chain scenario, however, this nonuniqueness does not affect global compatibility. The observed $AB$ marginal already fixes the action of the first pseudo-channel on the operators entering the sequential PDO representation. Consequently, the global consistency problem can be reduced to determining whether a second pseudo-channel compatible with $R_{BC}$ also reproduces the remaining correlations. This leads to the following criterion.

\begin{theorem}[Pseudo-Markov chain compatibility]
\label{thm:closure}
Let $R_{ABC}$ be a three-qubit PDO with Pauli data $(\boldsymbol u,\boldsymbol v,\boldsymbol w,M,N,K,G)$. Then $R_{ABC}$ is pseudo-Markov compatible with the chain $A\to B\to C$ if and only if there exists $\boldsymbol t\in\mathbb R^3$ such that
\begin{equation}
\boldsymbol w=N^{\mathsf T}\boldsymbol v+(1-\|\boldsymbol v\|^2)\boldsymbol t,
\label{eq:w}
\end{equation}
\begin{equation}
K=MN+(\boldsymbol u-M\boldsymbol v)\boldsymbol t^{\mathsf T},
\label{eq:K}
\end{equation}
and
\begin{equation}
G_{abc}=u_aN_{bc}+(M_{ab}-u_av_b)t_c,
\label{eq:G}
\end{equation}
for all $a,b,c=1,2,3$.
\end{theorem}

The details of the derivation are provided in Appendix~\ref{app:proof}. Theorem~\ref{thm:closure} characterizes the global consistency constraints imposed by the chain structure on the observed Pauli data. Equation~\eqref{eq:w} ensures compatibility with the $BC$ marginal, while Eqs.~\eqref{eq:K} and \eqref{eq:G} constrain the nonadjacent $AC$ correlations and the genuine three-body correlations, respectively. Crucially, all three conditions must be satisfied by the same translation vector $\boldsymbol t$ of the second pseudo-channel. Since these conditions are affine in $\boldsymbol t$, pseudo-Markov chain compatibility reduces to a linear feasibility problem in three real variables. When the intermediate marginal is full rank, the remaining parameters are uniquely determined, as stated in the following corollary.

\begin{corollary}
\label{cor:full-rank-B}
Let $R_{ABC}$ satisfy the assumptions of Theorem~\ref{thm:closure}, and suppose that $\rho_B=(I+\boldsymbol v\cdot\boldsymbol\sigma)/2$ is full rank, i.e., $\|\boldsymbol v\|<1$. Then the translation vector of the second pseudo-channel $\mathcal E_{C|B}$ is uniquely determined by the marginal $BC$ as
\begin{equation}
\boldsymbol t=\frac{\boldsymbol w-N^{\mathsf T}\boldsymbol v}{1-\|\boldsymbol v\|^2}.
\label{eq:chain_t_fullrank}
\end{equation}
Consequently, $R_{ABC}$ is pseudo-Markov compatible with the chain $A\to B\to C$ if and only if this vector satisfies Eqs.~\eqref{eq:K} and \eqref{eq:G}.
\end{corollary}

When the intermediate marginal $\rho_B$ is rank deficient, $\|\boldsymbol v\|=1$, and Eq.~\eqref{eq:w} reduces to $\boldsymbol w=N^{\mathsf T}\boldsymbol v$. In this case, the $BC$ marginal does not determine the translation vector $\boldsymbol t$, and chain compatibility requires this equality together with a common solution to Eqs.~\eqref{eq:K} and \eqref{eq:G}. Whether the compatible pseudo-channels can be chosen to be completely positive, thereby yielding a physical Markov realization, is examined in Sec.~\ref{sec:physicality}.

\section{Fork Causal Compatibility}
\label{sec:fork}

We next consider the fork structure $\mathsf G_{\mathrm f}:A\to B,\ A\to C$, where $A$ is the common cause of $B$ and $C$. Unlike the chain, the two child events have no direct causal influence on one another, although they may still be correlated through their common parent. The fork hypothesis requires these correlations to be consistent with two outgoing causal mechanisms satisfying the quantum Markov condition. To formulate this requirement, consider a qubit initially prepared in state $\rho_A$, which is measured at $A$ and subsequently transformed into two output qubits by a joint CPTP map $\Gamma_{BC|A}$. The output qubits are then measured at $B$ and $C$, respectively, according to the Pauli measurement scheme introduced in Sec.~\ref{sec:framework}. The resulting PDO admits the following representation.

\begin{proposition}
\label{prop:fork_PDO_representation}
Let $\rho_A$ be the initial state at $A$, and let $\Gamma_{BC|A}$ be a joint CPTP map from $A$ to $BC$. The corresponding three-qubit PDO is
\begin{equation}
R_{ABC}
=\frac14\sum_{a=0}^{3}
\sigma_a\otimes
\Gamma_{BC|A}\left(\{\rho_A,\sigma_a\}\right).
\label{eq:pseudo_fork_PDO}
\end{equation}
\end{proposition}

The proof is given in Appendix~\ref{app:fork_PDO_representation}. Equation~\eqref{eq:pseudo_fork_PDO} describes the correlations generated by a general joint channel from $A$ to $BC$. However, the existence of such a channel alone does not establish compatibility with a quantum Markov fork. As discussed in Sec.~\ref{sec:framework}, the Markov condition requires the joint mechanism to factorize into conditional channel operators associated with the two outgoing edges. Following Allen \textit{et al.}~\cite{Allen2017}, this factorization takes the form
\begin{equation}
J_{BC|A}=J_{B|A}J_{C|A},
\label{eq:fork_Allen_factorization}
\end{equation}
where $J_{B|A}$ and $J_{C|A}$ act on a common input space and are understood to be embedded in the joint operator space. Since the conditional Choi operators are Hermitian, their product must satisfy
\begin{equation}
[J_{B|A},J_{C|A}]=0.
\end{equation}
Thus, compatibility with a Markov fork requires more than the separate realization of the two parent--child marginals: the outgoing mechanisms must also admit a common factorized realization.

We now seek to express the pseudo-Markov fork compatibility conditions directly in terms of the observed Pauli data. Let $\boldsymbol p,\boldsymbol q\in\mathbb R^3$ denote the translation vectors of the outgoing HPTP pseudo-channels $\mathcal E_{B|A}$ and $\mathcal E_{C|A}$, respectively, whose conditional Choi operators satisfy the commutativity and factorization requirements discussed above. Applying Lemma~\ref{lem:pseudo-channel-reconstruction} to $R_{AB}$ and $R_{AC}$ allows the linear parts of both pseudo-channels to be expressed in terms of the observed marginals and these translation vectors. Defining $d_A:=1-\|\boldsymbol u\|^2=4\det\rho_A$ and denoting the $b$th and $c$th columns of $M$ and $K$ by $M_{:b}$ and $K_{:c}$, respectively, we obtain the following characterization.

\begin{theorem}[Pseudo-Markov fork compatibility]
\label{thm:pseudo_fork}
Let $R_{ABC}$ be a three-qubit PDO with Pauli data $(\boldsymbol u,\boldsymbol v,\boldsymbol w,M,N,K,G)$. Then $R_{ABC}$ is pseudo-Markov compatible with the fork $A\to B,\ A\to C$ if and only if there exists $\boldsymbol p,\boldsymbol q\in\mathbb R^3$ such that
\begin{equation}
d_A\boldsymbol p=\boldsymbol v-M^{\mathsf T}\boldsymbol u,~~d_A\boldsymbol q=\boldsymbol w-K^{\mathsf T}\boldsymbol u,
\label{eq:fork_F0}
\end{equation}
together with
\begin{equation}
\begin{aligned}
M_{:b}\times K_{:c}-q_c M_{:b}\times\boldsymbol u-p_b\boldsymbol u\times K_{:c}=\boldsymbol 0
\end{aligned}
\label{eq:fork_F1}
\end{equation}
for all $b,c=1,2,3$,
\begin{equation}
N-M^{\mathsf T}K=(\boldsymbol v-M^{\mathsf T}\boldsymbol u)\boldsymbol q^{\mathsf T},
\label{eq:fork_F2}
\end{equation}
and
\begin{equation}
\begin{aligned}
G_{abc}={}&u_aN_{bc}+(M_{ab}-u_av_b)q_c\\
&+p_b(K_{ac}-u_aw_c)
\end{aligned}
\label{eq:fork_F3}
\end{equation}
for all $a,b,c=1,2,3$. 
\end{theorem}

The proof is given in Appendix~\ref{app:fork_proof}. Theorem~\ref{thm:pseudo_fork} provides an observable-data criterion for the common-cause factorization required by a pseudo-Markov fork. Equation~\eqref{eq:fork_F0} ensures compatibility with the two parent--child marginals, while Eq.~\eqref{eq:fork_F1} enforces commutativity of the outgoing conditional operators. Equations~\eqref{eq:fork_F2} and \eqref{eq:fork_F3} require the same mechanisms to reproduce the child--child and genuine three-body correlations, respectively. Remarkably, despite the product of conditional operators in Eq.~\eqref{eq:fork_Allen_factorization}, all these conditions are affine in $\boldsymbol p$ and $\boldsymbol q$. Pseudo-Markov fork compatibility therefore reduces to a linear feasibility problem in six real variables. When the parent marginal is full rank, these variables are uniquely determined by the parent--child marginals.

\begin{corollary}
\label{cor:fork_full_rank}
Let $R_{ABC}$ satisfy the assumptions of Theorem~\ref{thm:pseudo_fork}, and suppose that $\rho_A$ is full ranked, i.e., $\|\boldsymbol u\|<1$. Then the translation vectors of the two outgoing pseudo-channels are uniquely determined by
\begin{equation}
\boldsymbol p=\frac{\boldsymbol v-M^{\mathsf T}\boldsymbol u}{d_A},~~\boldsymbol q=\frac{\boldsymbol w-K^{\mathsf T}\boldsymbol u}{d_A}.
\label{eq:fork_full_rank_pq}
\end{equation}
Consequently, $R_{ABC}$ is pseudo-Markov compatible with the fork $A\to B,\ A\to C$ if and only if these vectors satisfy Eqs.~\eqref{eq:fork_F1}--\eqref{eq:fork_F3}.
\end{corollary}

By contrast, when the parent marginal is rank deficient, $\|\boldsymbol u\|=1$, and Eq.~\eqref{eq:fork_F0} reduces to $\boldsymbol v=M^{\mathsf T}\boldsymbol u$ and $\boldsymbol w=K^{\mathsf T}\boldsymbol u$. Equation~\eqref{eq:fork_F2} then becomes $N=M^{\mathsf T}K$. Provided these data-only conditions hold, pseudo-Markov compatibility is equivalent to the existence of $\boldsymbol p,\boldsymbol q\in\mathbb R^3$ satisfying Eqs.~\eqref{eq:fork_F1} and \eqref{eq:fork_F3}. Thus, even when the parent--child marginals do not uniquely determine the outgoing mechanisms, the full three-event data can still constrain their admissible common-cause realizations.

\section{Collider Causal Compatibility}
\label{sec:collider}
We finally consider the collider structure $\mathsf G_{\mathrm{col}}:A\to C\leftarrow B$, where two independent root nodes $A$ and $B$ jointly influence the common child $C$. Unlike the chain and fork structures, the collider has two independent preparations and a single nontrivial causal mechanism acting on the joint parent system $AB$. The independence of the roots constrains their mutual correlations, while their joint influence on $C$ may generate correlations involving all three events. To characterize this structure, consider two qubits independently prepared in states $\rho_A$ and $\rho_B$, respectively. Following Pauli measurements at $A$ and $B$, the two systems are processed by a joint CPTP map $\mathcal E_{C|AB}$, and the resulting qubit is measured at $C$. The corresponding PDO has the following representation.

\begin{proposition}
\label{prop:collider_PDO}
Let $\rho_A$ and $\rho_B$ be the independently prepared qubit states at the root nodes $A$ and $B$, respectively, and let $\mathcal E_{C|AB}$ be a CPTP map from $AB$ to $C$. The corresponding three-qubit PDO is
\begin{equation}
R_{ABC}=\frac{1}{16}\sum_{a,b=0}^{3}\sigma_a\otimes\sigma_b\otimes\mathcal E_{C|AB}\left(\{\rho_A,\sigma_a\}\otimes\{\rho_B,\sigma_b\}\right).
\label{eq:collider_PDO}
\end{equation}
\end{proposition}

The proof is provided in Appendix~\ref{app:collider_representation}. To address the inverse problem, we retain the independent root preparations in Eq.~\eqref{eq:collider_PDO} while relaxing the joint CPTP channel to an HPTP pseudo-channel. The task is then to determine whether the observed Pauli correlations admit such a representation. We show that this requirement can be formulated as a linear feasibility problem.

To express the action of the joint pseudo-channel in terms of the observed Pauli data, we first note that the anticommutators arising from the Lüders updates at the root nodes are linear combinations of Pauli operators. For a qubit state $\rho=(I+\boldsymbol r\cdot\boldsymbol\sigma)/2$, these combinations can be encoded by the real symmetric matrix
\begin{equation}
H(\boldsymbol r)=
\begin{pmatrix}
1 & \boldsymbol r^{\mathsf T}\\
\boldsymbol r & I_3
\end{pmatrix},
\label{eq:collider_H}
\end{equation}
such that $\{\rho,\sigma_a\}=\sum_{\mu=0}^{3}H(\boldsymbol r)_{\mu a}\sigma_\mu$. Since $\det H(\boldsymbol r)=1-\|\boldsymbol r\|^2$, this matrix is invertible precisely when $\rho$ is full rank. We parameterize the joint HPTP pseudo-channel by its real Pauli-transfer coefficients,
\begin{equation}
\mathcal E_{C|AB}(\sigma_\mu\otimes\sigma_\nu)=2\sum_{c=0}^{3}\ell_{c|\mu\nu}\sigma_c,
\label{eq:collider_Pauli_transfer}
\end{equation}
and collect them into matrices $L^{(c)}=[\ell_{c|\mu\nu}]_{\mu,\nu=0}^{3}\in\mathbb R^{4\times4}$. Trace preservation fixes $L^{(0)}=E_{00}:=\boldsymbol e_0\boldsymbol e_0^{\mathsf T}$, where $\boldsymbol e_0=(1,0,0,0)^{\mathsf T}$, leaving three real matrices $L^{(1)},L^{(2)},L^{(3)}$ to characterize the joint pseudo-channel.

We similarly arrange the observed Pauli correlators into matrices $\mathcal R^{(c)}=[r_{abc}]_{a,b=0}^{3}$ for $c=0,1,2,3$. In terms of the Pauli data introduced in Sec.~\ref{sec:framework}, these matrices take the form
\begin{equation}
\mathcal R^{(0)}=
\begin{pmatrix}
1 & \boldsymbol v^{\mathsf T}\\
\boldsymbol u & M
\end{pmatrix},
\label{eq:collider_R0}
\end{equation}
and, for $c=1,2,3$,
\begin{equation}
\mathcal R^{(c)}=
\begin{pmatrix}
w_c & N_{:c}^{\mathsf T}\\
K_{:c} & G^{(c)}
\end{pmatrix},
\label{eq:collider_Rc}
\end{equation}
where $G^{(c)}_{ab}=G_{abc}$. The matrices $H(\boldsymbol u)$ and $H(\boldsymbol v)$ encode the two root preparations and their measurement updates, while $L^{(c)}$ describes the subsequent joint mechanism. Their relation to the observed correlations yields the following compatibility criterion.

\begin{theorem}[Pseudo-Markov collider compatibility]
\label{thm:collider}
Let $R_{ABC}$ be a three-qubit PDO with Pauli data $(\boldsymbol u,\boldsymbol v,\boldsymbol w, M,N,K,G)$. Then $R_{ABC}$ is pseudo-Markov compatible with the collider $A\to C\leftarrow B$ if and only if
\begin{equation}
M=\boldsymbol u\boldsymbol v^{\mathsf T},
\label{eq:collider_root_independence}
\end{equation}
and there exist real matrices $L^{(1)},L^{(2)},L^{(3)}\in\mathbb R^{4\times4}$ satisfying
\begin{equation}
\mathcal R^{(c)}=H(\boldsymbol u)^{\mathsf T}L^{(c)}H(\boldsymbol v),~~ c=1,2,3.
\label{eq:collider_feasibility}
\end{equation}
\end{theorem}

The proof is given in Appendix~\ref{app:collider_proof}. Theorem~\ref{thm:collider} separates the compatibility problem into two requirements: Eq.~\eqref{eq:collider_root_independence} imposes the independence of the root systems, while Eq.~\eqref{eq:collider_feasibility} requires the observed correlations involving $C$ to be reproduced by a single joint HPTP map. Since the latter equations are linear in the unknown Pauli-transfer coefficients, pseudo-Markov collider compatibility reduces to a linear feasibility problem. Moreover, the compatible joint pseudo-channels can be explicitly characterized using the Moore--Penrose inverse, as shown in the following corollary.

\begin{corollary}
\label{cor:collider_reconstruction}
Suppose that $R_{ABC}$ is pseudo-Markov compatible with $A\to C\leftarrow B$. Then all compatible Pauli-transfer matrices are given by
\begin{equation}
\begin{aligned}
L^{(c)}={}&H(\boldsymbol u)^+\mathcal R^{(c)}H(\boldsymbol v)^++Y_c\\
&-H(\boldsymbol u)^+H(\boldsymbol u)Y_cH(\boldsymbol v)H(\boldsymbol v)^+,
\end{aligned}
\label{eq:collider_general_reconstruction}
\end{equation}
for $c=1,2,3$, where $L^{(0)}=E_{00}$ and $Y_c\in\mathbb R^{4\times4}$ is arbitrary.
\end{corollary}

The proof is provided in Appendix~\ref{app:collider_reconstruction_proof}. If both root states are full rank, the matrices $H(\boldsymbol u)$ and $H(\boldsymbol v)$ are invertible, and the joint pseudo-channel is uniquely determined by
\begin{equation}
L^{(c)}=H(\boldsymbol u)^{-1}\mathcal R^{(c)}H(\boldsymbol v)^{-1},\qquad c=1,2,3.
\label{eq:collider_fullrank_reconstruction}
\end{equation}
In this case, Eq.~\eqref{eq:collider_feasibility} is always solvable, and pseudo-Markov collider compatibility reduces entirely to the root independence condition $M=\boldsymbol u\boldsymbol v^{\mathsf T}$. By contrast, if either root state is rank deficient, the reconstruction is generally nonunique and the observed correlations must satisfy additional consistency constraints. Defining the orthogonal projector onto the range of $H(\boldsymbol r)$ as $P(\boldsymbol r)=H(\boldsymbol r)H(\boldsymbol r)^+$, Eq.~\eqref{eq:collider_feasibility} is solvable if and only if
\begin{equation}
\mathcal R^{(c)}=P(\boldsymbol u)\mathcal R^{(c)}P(\boldsymbol v),\qquad c=1,2,3.
\label{eq:collider_projector}
\end{equation}

These projector conditions characterize the additional constraints imposed by rank-deficient root preparations on the observed Pauli data, while Eq.~\eqref{eq:collider_general_reconstruction} describes the remaining freedom in the compatible joint pseudo-channels.

\section{Physical Realizability and Its Quantification}
\label{sec:physicality}

The preceding sections establish necessary and sufficient conditions for pseudo-Markov compatibility with the chain, fork, and collider structures, without requiring the underlying HPTP mechanisms to be completely positive. We now investigate whether a compatible realization can be chosen to consist entirely of CPTP mechanisms and, when this is impossible, quantify the minimum departure from complete positivity among all compatible realizations.

For an HPTP map $\mathcal E$, let $\chi_{\mathcal E}=J_{\mathcal E}/d_{\mathrm{in}}$ denote its normalized Choi operator, where $d_{\mathrm{in}}$ is the input dimension. Since $\chi_{\mathcal E}$ is Hermitian with unit trace, we quantify its failure of positivity by the Choi negativity
\begin{equation}
\mathcal N(\chi_{\mathcal E}):=\frac{\|\chi_{\mathcal E}\|_1-1}{2},
\label{eq:choi_negativity}
\end{equation}
which equals the sum of the absolute values of its negative eigenvalues. For a trace-preserving map, this quantity vanishes if and only if the map is completely positive. Thus, the Choi negativity provides a direct measure of the nonphysicality of an individual pseudo-channel. For causal compatibility, however, the relevant question is not whether a particular reconstruction is completely positive, but whether at least one compatible realization consists entirely of physical mechanisms.

A given PDO may admit different pseudo-Markov realizations under the same causal structure. To account for this possibility, let $\mathcal C_{\mathsf G}^{\mathrm p}$ and $\mathcal C_{\mathsf G}$ denote the sets of pseudo-Markov-compatible and physically Markov-compatible PDOs for a causal graph $\mathsf G$, respectively, with $\mathcal C_{\mathsf G}\subseteq\mathcal C_{\mathsf G}^{\mathrm p}$. For $R\in\mathcal C_{\mathsf G}^{\mathrm p}$, let $\mathfrak R_{\mathsf G}^{\mathrm p}(R)$ denote the set of all pseudo-Markov realizations reproducing $R$, with physical root preparations. Each realization $\boldsymbol{\mathcal E}$ consists of the HPTP mechanisms associated with the nonroot nodes. We define the graph-dependent nonphysicality of $R$ as
\begin{equation}
f_{\mathsf G}(R):=\min_{\boldsymbol{\mathcal E}\in\mathfrak R_{\mathsf G}^{\mathrm p}(R)}\sum_{\mathcal E\in\boldsymbol{\mathcal E}}\mathcal N(\chi_{\mathcal E}).
\label{eq:graph_nonphysicality}
\end{equation}
Since Choi negativity is nonnegative and vanishes precisely for completely positive maps, $f_{\mathsf G}(R)=0$ if and only if $R\in\mathcal C_{\mathsf G}$. The minimum is attained in finite dimensions, as bounded Choi negativity implies bounded trace norm for each normalized Choi operator, while the compatibility constraints define a closed feasible set.

The usefulness of this formulation is that the algebraic compatibility conditions derived in the preceding sections are affine in the remaining pseudo-channel parameters. Complete positivity can therefore be incorporated through semidefinite constraints on their Choi operators. We illustrate this construction explicitly for the chain $\mathsf G_{\mathrm{ch}}:A\to B\to C$. Let $\boldsymbol t_1$ and $\boldsymbol t_2$ denote the translation vectors of the two pseudo-channels. By Lemma~\ref{lem:pseudo-channel-reconstruction}, their Bloch matrices are $T_1=M^{\mathsf T}-\boldsymbol t_1\boldsymbol u^{\mathsf T}$ and $T_2=N^{\mathsf T}-\boldsymbol t_2\boldsymbol v^{\mathsf T}$, respectively. For a qubit HPTP map with Bloch-affine parameters $(T,\boldsymbol t)$, the normalized Choi operator is
\begin{equation}
\chi(T,\boldsymbol t)=\frac14\left[I\otimes(I+\boldsymbol t\cdot\boldsymbol\sigma)+\sum_{a,b=1}^{3}T_{ba}\,\sigma_a^{\mathsf T}\otimes\sigma_b\right].
\label{eq:qubit_choi_physicality}
\end{equation}
Accordingly, we write $\chi_{B|A}(\boldsymbol t_1)=\chi(T_1,\boldsymbol t_1)$ and $\chi_{C|B}(\boldsymbol t_2)=\chi(T_2,\boldsymbol t_2)$, both of which depend affinely on their respective translation vectors.

The Choi negativity admits the semidefinite representation
\begin{equation}
\mathcal N(\chi)=\min_{\substack{P,Q\succeq0\\ \chi=P-Q}}\operatorname{Tr}Q,
\label{eq:negativity_sdp}
\end{equation}
where $P$ and $Q$ are auxiliary positive semidefinite operators. Combining this representation with the two-event reconstruction relations and the chain closure conditions of Theorem~\ref{thm:closure}, we obtain
\begin{equation}
\begin{aligned}
f_{\mathsf G_{\mathrm{ch}}}(R)
=\min_{\substack{\boldsymbol t_1,\boldsymbol t_2\\ P_1,Q_1,P_2,Q_2}}
~&\operatorname{Tr}Q_1+\operatorname{Tr}Q_2\\
\mathrm{s.t.}\quad
&(1-\|\boldsymbol u\|^2)\boldsymbol t_1
=\boldsymbol v-M^{\mathsf T}\boldsymbol u,\\
&\boldsymbol w=N^{\mathsf T}\boldsymbol v
+(1-\|\boldsymbol v\|^2)\boldsymbol t_2,\\
&K=MN+(\boldsymbol u-M\boldsymbol v)
\boldsymbol t_2^{\mathsf T},\\
&G_{abc}=u_aN_{bc}+(M_{ab}-u_av_b)t_{2c},\\
&\chi_{B|A}(\boldsymbol t_1)=P_1-Q_1,\\
&\chi_{C|B}(\boldsymbol t_2)=P_2-Q_2,\\
&P_1,Q_1,P_2,Q_2\succeq0.
\end{aligned}
\label{eq:chain_nonphysicality_sdp}
\end{equation}
Here, the tensor constraint holds for all $a,b,c=1,2,3$. For fixed Pauli data, all equality constraints are affine in the optimization variables, while the remaining constraints are linear matrix inequalities. Equation~\eqref{eq:chain_nonphysicality_sdp} is therefore a semidefinite program whose optimal value gives the minimum total Choi negativity required by a sequential pseudo-Markov explanation. In particular, the existence of a physical Markov-chain realization is equivalent to a zero optimal value.

The fork structure $\mathsf G_{\mathrm f}:A\to B,\ A\to C$ can be treated in the same way. The two outgoing pseudo-channels have Bloch-affine parameters $(M^{\mathsf T}-\boldsymbol p\boldsymbol u^{\mathsf T},\boldsymbol p)$ and $(K^{\mathsf T}-\boldsymbol q\boldsymbol u^{\mathsf T},\boldsymbol q)$, respectively. Their normalized Choi operators are affine in $\boldsymbol p$ and $\boldsymbol q$, while Eqs.~\eqref{eq:fork_F0}--\eqref{eq:fork_F3} encode the parent--child compatibility, commutativity, and global consistency requirements. Minimizing $\mathcal N(\chi_{B|A})+\mathcal N(\chi_{C|A})$ subject to these affine constraints consequently yields an SDP for $f_{\mathsf G_{\mathrm f}}(R)$.

For the collider $\mathsf G_{\mathrm{col}}:A\to C\leftarrow B$, physical realizability concerns a single joint-input pseudo-channel $\mathcal E_{C|AB}$. Using the Pauli-transfer coefficients introduced in Eq.~\eqref{eq:collider_Pauli_transfer}, its normalized Choi operator takes the form
\begin{equation}
\chi_{C|AB}=\frac18\sum_{\mu,\nu,c=0}^{3}\ell_{c|\mu\nu}\,\sigma_\mu^{\mathsf T}\otimes\sigma_\nu^{\mathsf T}\otimes\sigma_c,
\label{eq:collider_choi_physicality}
\end{equation}
where trace preservation fixes $L^{(0)}=E_{00}$. Since $\chi_{C|AB}$ depends affinely on the remaining transfer matrices $L^{(1)},L^{(2)},L^{(3)}$, minimizing its Choi negativity subject to the root independence condition and Eq.~\eqref{eq:collider_feasibility} likewise gives an SDP for $f_{\mathsf G_{\mathrm{col}}}(R)$. Hence, the three causal structures, despite imposing different algebraic constraints on the observed correlations, admit a unified treatment of physical realizability through the minimization of Choi negativity over their respective compatible pseudo-Markov mechanisms.

\section{Examples}
\label{sec:examples}
We illustrate the preceding results by examining the causal compatibility of a two-parameter family of three-qubit PDOs. We first compare the physical compatibility regions of the chain, fork, and collider structures within the same parameter space. We then select a representative PDO to demonstrate how the algebraic compatibility criteria and the graph-dependent nonphysicality measure can be used to assess its compatibility with different causal structures.

Consider the family
\begin{equation}
\begin{aligned}
R_r(x,y)=\frac18\Big[&III+r\sigma_xII+x\sigma_z\sigma_zI+yI\sigma_z\sigma_z\\
&+\frac{x+y}{2}\sigma_zI\sigma_z+ry\sigma_x\sigma_z\sigma_z\Big],
\end{aligned}
\label{eq:landscape_family}
\end{equation}
where tensor products between single-qubit operators are implicit, $r\in(0,1)$ is fixed, and $(x,y)\in[-1,1]^2$. We denote this family by $\mathscr R_r=\{R_r(x,y):(x,y)\in[-1,1]^2\}$. Its Pauli data are $\boldsymbol u=r\boldsymbol e_x$, $\boldsymbol v=\boldsymbol w=\boldsymbol0$, $M=xP_z$, $N=yP_z$, $K=(x+y)P_z/2$, and $G_{133}=ry$, with all other three-body coefficients vanishing. Here, $P_z=\boldsymbol e_z\boldsymbol e_z^{\mathsf T}$, and $\boldsymbol e_x$ and $\boldsymbol e_z$ are the standard unit vectors along the $x$ and $z$ axes of Bloch space, respectively.

\begin{figure}[t]
\centering
\includegraphics[width=0.9\columnwidth]
{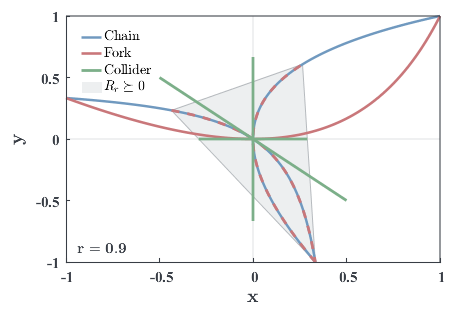}
\caption{
Physical causal-compatibility landscape of the PDO family $R_r(x,y)$ in Eq.~\eqref{eq:landscape_family} for $r=0.9$. Blue curves represent the union over all six chain orderings, red curves the union over all three choices of common parent for the fork, and green segments the union over all three choices of common child for the collider. Red dashed portions overlapping the blue curves indicate PDOs admitting both chain and fork explanations. The gray region corresponds to $R_r(x,y)\succeq0$, where the Pauli correlations admit a spatial quantum-state realization.
}
\label{fig:compatibility_landscape}
\end{figure}

For each topology $\tau\in\{\mathrm{chain},\mathrm{fork},\mathrm{coll}\}$, let $\mathcal C_\tau$ and $\mathcal C_\tau^{\mathrm p}$ denote the unions of the physical and pseudo-Markov compatibility sets over all node assignments of that topology, respectively. Applying the criteria derived in the preceding sections to $\mathscr R_r$ gives the physical compatibility landscape shown in Fig.~\ref{fig:compatibility_landscape}. The three causal classes occupy distinct subsets of the parameter space, with intersections corresponding to PDOs admitting multiple Markov causal explanations. At the same time, none of the three classes is contained in either of the others, indicating that the observed correlations can distinguish certain causal structures even when a unique explanation is not available.

To illustrate the compatibility analysis for a fixed set of observed correlations, we choose $r=0.9$, $x=-0.75$, and $y=0.3$, and denote the corresponding PDO by $R_*$. We examine whether this PDO admits chain, fork, or collider explanations, distinguishing structures excluded by the algebraic compatibility conditions from those whose compatible mechanisms fail to satisfy complete positivity.

First, consider the chain $A\to B\to C$. Since $\boldsymbol v=\boldsymbol w=\boldsymbol{0}$ and $M^{\mathsf T}\boldsymbol u=\boldsymbol{0}$, the two adjacent marginals uniquely determine compatible HPTP pseudo-channels with Bloch-affine parameters $(T_1,\boldsymbol t_1)=(xP_z,\boldsymbol{0})$ and $(T_2,\boldsymbol t_2)=(yP_z,\boldsymbol{0})$. To establish global compatibility, we apply the closure conditions of Theorem~\ref{thm:closure}. Equation~\eqref{eq:w} is satisfied because $\boldsymbol v=\boldsymbol w=\boldsymbol t_2=\boldsymbol0$, while Eq.~\eqref{eq:K} requires $K=MN$, which holds since $(x+y)/2=xy$. Moreover, Eq.~\eqref{eq:G} reduces to $G_{abc}=u_aN_{bc}$, reproducing the only nonzero three-body coefficient $G_{133}=ry$. Thus, the complete Pauli data satisfy the chain closure conditions. Both reconstructed pseudo-channels are CPTP, since their Bloch matrices have the form $\lambda P_z$ with $|\lambda|\le1$ and vanishing translation vectors. Consequently, $R_*$ admits a physical Markov-chain realization, giving $f_{A\to B\to C}(R_*)=0$.

Next, consider the fork $B\to A,\ B\to C$. To apply Theorem~\ref{thm:pseudo_fork}, we relabel the events in the order $(B,A,C)$. The common parent has Bloch vector $\boldsymbol0$, the two parent--child correlation matrices are $xP_z$ and $yP_z$, and the child--child correlation matrix is $(x+y)P_z/2$. Since the parent state $\rho_B=I/2$ is full rank, Eq.~\eqref{eq:fork_F0} uniquely determines the translation vectors $\boldsymbol p=r\boldsymbol e_x$ and $\boldsymbol q=\boldsymbol0$. The corresponding outgoing pseudo-channels are
\begin{equation*}
(T_{A|B},\boldsymbol t_{A|B})=(xP_z,r\boldsymbol e_x),~
(T_{C|B},\boldsymbol t_{C|B})=(yP_z,\boldsymbol0).
\label{eq:example_fork_channels}
\end{equation*}
We verify their pseudo-Markov compatibility using the remaining conditions of Theorem~\ref{thm:pseudo_fork}. Since the parent Bloch vector vanishes and both parent--child correlation matrices are proportional to $P_z$, the commutativity condition \eqref{eq:fork_F1} holds. Equation~\eqref{eq:fork_F2} reduces to $(x+y)/2=xy$, which is satisfied by the selected parameters. Finally, Eq.~\eqref{eq:fork_F3}, expressed in the reordered variables $(B,A,C)$, reproduces the only nonzero three-body coefficient $G_{133}=ry$ through the term $p_1y$, while all other coefficients vanish as required. Hence, $R_*$ is pseudo-Markov compatible with this fork assignment.

We next examine the complete positivity of compatible fork mechanisms. The normalized Choi operator of $\mathcal E_{A|B}$ has negative eigenvalues $\lambda_\pm=(1-\sqrt{r^2+x^2})/4$ with multiplicity two, and then it is not completely positive. By contrast, $\mathcal E_{C|B}$ is CPTP because $|y|\le1$. Moreover, the full-rank parent marginal uniquely determines both outgoing pseudo-channels, so no alternative compatible realization can eliminate this violation of complete positivity. The minimum total Choi negativity is therefore $f_{B\to\{A,C\}}(R_*)=(\sqrt{r^2+x^2}-1)/2\approx0.08577>0$. Thus, although $R_*$ admits a pseudo-Markov fork explanation with common parent $B$, no fork assignment provides a physical Markov realization.

Finally, consider collider structures. All three node assignments can be excluded directly using the root-independence condition of Theorem~\ref{thm:collider}, without reconstructing a joint-input pseudo-channel. Depending on the choice of the common child, this condition requires $M=\boldsymbol u\boldsymbol v^{\mathsf T}$, $K=\boldsymbol u\boldsymbol w^{\mathsf T}$, or $N=\boldsymbol v\boldsymbol w^{\mathsf T}$. For $R_*$, we have $\boldsymbol v=\boldsymbol w=\boldsymbol0$, whereas $M$, $K$, and $N$ are all nonzero. Hence, none of these independence conditions is satisfied, and every collider assignment is excluded at the pseudo-Markov level.

Combining these results, we obtain $R_*\in\mathcal C_{\mathrm{chain}}\cap\mathcal C_{\mathrm{fork}}^{\mathrm p}$ and $R_*\notin\mathcal C_{\mathrm{fork}}\cup\mathcal C_{\mathrm{coll}}^{\mathrm p}$. This example illustrates how the algebraic compatibility criteria and the graph-dependent nonphysicality measure can be used together to distinguish different causal structures from the same observed correlations.

\section{Discussion}

We have characterized the causal compatibility of three-qubit PDOs with the chain, fork, and collider structures directly from observed Pauli correlations, without requiring an intervention-complete description of the underlying process. By separating the algebraic constraints of quantum Markov structures from the complete positivity of their local mechanisms, we obtained necessary and sufficient conditions for pseudo-Markov compatibility with each causal graph. These conditions reduce to linear feasibility problems, providing a direct way to assess whether the observed correlations admit a candidate Markov causal explanation.

To determine physical realizability, we introduced a graph-dependent nonphysicality measure defined by minimizing the total Choi negativity over all compatible pseudo-Markov mechanisms. This measure vanishes if and only if the observed PDO admits a physical Markov realization of the specified causal graph, and its minimization can be formulated as a semidefinite program. The two-parameter PDO family illustrates how different causal structures can admit overlapping physical compatibility regions, while the analysis of a representative PDO demonstrates how algebraic compatibility tests and the nonphysicality measure can yield different conclusions for different causal structures from the same observed correlations. These results establish a systematic approach to assessing candidate quantum Markov explanations using experimentally accessible correlation data.

Our analysis is restricted to three-qubit events, the specified Pauli measurement scheme, and quantum Markov models associated with the candidate causal graphs. Extending the approach to larger networks and more general DAGs remains an important direction, particularly because the coupling between local mechanisms may complicate the elimination of unobserved parameters. For experimental applications, statistical uncertainties in measured Pauli correlations motivate robust formulations of the compatibility tests that account for finite measurement precision. More generally, causal ambiguities that persist under the fixed measurement scheme may require access to additional quantum instruments or controlled interventions. Investigating these extensions may help clarify the extent to which multipartite quantum causal structures can be characterized from restricted experimental data.

\begin{acknowledgments}
This work was supported by the National Natural Science Foundation of China under Grant No.~12501639, 12371458, Key Lab of Guangzhou for Quantum Precision Measurement under Grant No.~202201000010, and the Fundamental Research Funds for the Central Universities under Grant No.~21624351.
\end{acknowledgments}

\appendix

\section{Proof of Proposition~\ref{prop:R}}
\label{prop1:proof}

\begin{proof}
For a projective measurement of a Pauli observable $\sigma_a$, with $a\in\{1,2,3\}$, the L\"uders state update satisfies
\begin{equation}
\sum_{x=\pm1}x\,\Pi_{x|a}X\Pi_{x|a}=\frac12\{X,\sigma_a\}.
\label{eq:measurement_identity}
\end{equation}
This identity also holds for the identity setting $a=0$, where $\Pi_{+|0}=I$ and $\Pi_{-|0}=0$. Applying Eq.~\eqref{eq:measurement_identity} to the measurements at $A$ and $B$ in Eq.~\eqref{pro}, and using $\sum_{z=\pm1}z\Pi_{z|c}=\sigma_c$, we obtain
\begin{equation*}
r_{abc}=\frac14\operatorname{Tr}\left[\sigma_c\,\mathcal E_{C|B}\left(\left\{\mathcal E_{B|A}\left(\{\rho_A,\sigma_a\}\right),\sigma_b\right\}\right)\right].
\end{equation*}
For brevity, define $X_{ab}:=\mathcal E_{C|B}\left(\left\{\mathcal E_{B|A}(\{\rho_A,\sigma_a\}),\sigma_b\right\}\right)$. Using the Pauli completeness relation $\sum_{c=0}^{3}\operatorname{Tr}(X\sigma_c)\sigma_c=2X$, we find $\sum_{c=0}^{3}r_{abc}\sigma_c=X_{ab}/2$. Substituting this expression into the Pauli expansion \eqref{R_ABC} gives
\begin{align*}
R_{ABC}
&=\frac18\sum_{a,b=0}^{3}\sigma_a\otimes\sigma_b\otimes\left(\sum_{c=0}^{3}r_{abc}\sigma_c\right)\\
&=\frac1{16}\sum_{a,b=0}^{3}\sigma_a\otimes\sigma_b\otimes X_{ab},
\end{align*}
which establishes Eq.~\eqref{eq:physical-chain-pdo}.
\end{proof}

\section{Proof of Lemma~\ref{lem:pseudo-channel-reconstruction}}
\label{lemma2:proof}

\begin{proof}
A forward HPTP pseudo-channel $\mathcal E_{B|A}$ compatible with $R_{AB}$ reproduces the two-time PDO through
\begin{equation*}
R_{AB}=\frac14\sum_{\mu=0}^{3}\sigma_\mu\otimes\mathcal E_{B|A}\left(\{\rho_A,\sigma_\mu\}\right).
\end{equation*}
For a qubit HPTP map with Bloch-affine parameters $(T,\boldsymbol t)$, we have $\mathcal E_{B|A}(I)=I+\boldsymbol t\cdot\boldsymbol\sigma$ and $\mathcal E_{B|A}(\sigma_a)=\sum_{b=1}^{3}T_{ba}\sigma_b$. Since $\{\rho_A,\sigma_a\}=\sigma_a+u_aI$ for $a=1,2,3$, the two-body Pauli coefficients satisfy
\begin{equation*}
\begin{aligned}
M_{ab}&=\frac12\operatorname{Tr}\left[\sigma_b\mathcal E_{B|A}\left(\{\rho_A,\sigma_a\}\right)\right]\\
&=T_{ba}+u_at_b.
\end{aligned}
\end{equation*}
Furthermore, tracing over $A$ gives $\rho_B=\operatorname{Tr}_A R_{AB}=\mathcal E_{B|A}(\rho_A)$, and hence $\boldsymbol v=T\boldsymbol u+\boldsymbol t$. Combining these relations yields $T=M^{\mathsf T}-\boldsymbol t\boldsymbol u^{\mathsf T}$ and $(1-\|\boldsymbol u\|^2)\boldsymbol t=\boldsymbol v-M^{\mathsf T}\boldsymbol u$.

For $\|\boldsymbol u\|<1$, these relations uniquely determine $\boldsymbol t$ and $T$, yielding Eq.~\eqref{eq:two-time-full-rank}. In the rank-deficient case, $\|\boldsymbol u\|=1$, compatibility requires $\boldsymbol v=M^{\mathsf T}\boldsymbol u$. When this condition holds, $\boldsymbol t$ remains arbitrary, with $T=M^{\mathsf T}-\boldsymbol t\boldsymbol u^{\mathsf T}$, giving the affine family in Eq.~\eqref{eq:two-time-rank-deficient-family}.

Conversely, any real parameters $(T,\boldsymbol t)$ satisfying Eq.~\eqref{eq:two-time-compatibility} define an HPTP map through the specified action on the Pauli basis. Indeed, this action preserves Hermiticity and trace, while the resulting two-time PDO reproduces the prescribed local Bloch vectors and two-body Pauli coefficients. Since these coefficients completely determine $R_{AB}$, the constructed map is compatible with the given PDO. This establishes the claimed characterization.
\end{proof}

\section{Proof of Theorem~\ref{thm:closure}}
\label{app:proof}
\begin{proof}
We first establish necessity. Suppose that $R_{ABC}$ admits a pseudo-Markov chain representation $A\to B\to C$ with HPTP pseudo-channels $\mathcal E_{B|A}$ and $\mathcal E_{C|B}$. Compatibility of the first pseudo-channel with $R_{AB}$ implies, by Lemma~\ref{lem:pseudo-channel-reconstruction},
\begin{equation*}
\mathcal E_{B|A}(\{\rho_A,\sigma_a\})=u_a I+\sum_{b=1}^{3}M_{ab}\sigma_b,
\end{equation*}
for $a=1,2,3$. In particular, the first causal segment satisfies the two-event compatibility condition stated in the theorem.

Let $(T,\boldsymbol t)$ denote the Bloch-affine parameters of the second pseudo-channel $\mathcal E_{C|B}$. Applying Lemma~\ref{lem:pseudo-channel-reconstruction}
to $R_{BC}$ gives $N=T^{\mathsf T} +\boldsymbol v\boldsymbol t^{\mathsf T}$ and $\boldsymbol w=T\boldsymbol v+\boldsymbol t$. Thus $T_{cb}=N_{bc}-v_bt_c$, and substitution into the relation for $\boldsymbol w$ immediately yields Eq.~\eqref{eq:w}.

We next consider correlations involving the earlier event $A$. When no measurement is performed at $B$, the measurement scheme specified above yields
\begin{equation*}
\begin{aligned}
K_{ac}
&=\frac12\operatorname{Tr}\left[\sigma_c\,\mathcal E_{C|B}\left(\mathcal E_{B|A}
(\{\rho_A,\sigma_a\})\right)\right]\\
&=u_at_c+\sum_{b=1}^{3}M_{ab}T_{cb}.
\end{aligned}
\end{equation*}
Using $T_{cb}=N_{bc}-v_bt_c$, we obtain Eq.~\eqref{eq:K}.

For the genuine three-body correlations, the Pauli anticommutation relations give
\begin{equation*}
\left\{\mathcal E_{B|A}(\{\rho_A,\sigma_a\}),\sigma_b\right\}=2u_a\sigma_b+2M_{ab}I.
\end{equation*}
Consequently, Eq.~\eqref{eq:physical-chain-pdo} implies
\begin{equation*}
\begin{aligned}
G_{abc}
&=\frac14\operatorname{Tr}\left[\sigma_c\,\mathcal E_{C|B}\left(2u_a\sigma_b+2M_{ab}I\right)\right]\\
&=u_aT_{cb}+M_{ab}t_c\\
&=u_aN_{bc}+(M_{ab}-u_av_b)t_c,
\end{aligned}
\end{equation*}
which is Eq.~\eqref{eq:G}. This establishes the necessity of the closure conditions.

Conversely, suppose that the first-segment compatibility condition holds and that there exists $\boldsymbol t\in\mathbb R^3$ satisfying Eqs.~\eqref{eq:w}--\eqref{eq:G}. By Lemma~\ref{lem:pseudo-channel-reconstruction}, we can choose an HPTP pseudo-channel $\mathcal E_{B|A}$ compatible with $R_{AB}$. Define the second map on the Pauli basis by
\begin{equation*}
\begin{aligned}
\mathcal E_{C|B}(I)
&=I+\boldsymbol t\cdot\boldsymbol\sigma,\\
\mathcal E_{C|B}(\sigma_b)
&=\sum_{c=1}^{3}(N_{bc}-v_bt_c)\sigma_c.
\end{aligned}
\end{equation*}
The reality of the coefficients ensures Hermiticity preservation, while $\operatorname{Tr}\mathcal E_{C|B}(I)=2$ and $\operatorname{Tr}\mathcal E_{C|B}(\sigma_b)=0$ establish trace preservation. Hence $\mathcal E_{C|B}$ is HPTP, with Bloch matrix $T_{cb}=N_{bc}-v_bt_c$.

Let $\widetilde R_{ABC}$ be the PDO generated by $\rho_A$, $\mathcal E_{B|A}$, and $\mathcal E_{C|B}$ through Eq.~\eqref{eq:physical-chain-pdo}. By construction, its $AB$ marginal agrees with $R_{AB}$, and its $BC$ correlation matrix is $N$. Equation~\eqref{eq:w} ensures that the Bloch vector at $C$ is $\boldsymbol w$. Moreover, the expressions for the $AC$ and three-body coefficients derived above, together with Eqs.~\eqref{eq:K} and \eqref{eq:G}, show that the reconstructed correlations coincide with $K$ and $G$, respectively.

Thus $\widetilde R_{ABC}$ reproduces all the Pauli coefficients of $R_{ABC}$. Since the three-qubit Pauli products form an operator basis, we have $\widetilde R_{ABC}=R_{ABC}$. The constructed pseudo-channels therefore provide the required pseudo-Markov chain representation, establishing sufficiency.
\end{proof}

\section{Proof of Proposition~\ref{prop:fork_PDO_representation}}
\label{app:fork_PDO_representation}
\begin{proof}
Consider the fork structure $A\to B,\ A\to C$ with common-parent state $\rho_A$ and joint HPTP pseudo-channel $\Gamma_{BC|A}$. Applying Eq.~\eqref{eq:measurement_identity} at $A$ and evaluating the expectation values of the Pauli observables at $B$ and $C$, we obtain
\begin{equation*}
r_{abc}=\frac12\operatorname{Tr}\left[(\sigma_b\otimes\sigma_c)\Gamma_{BC|A}\left(\{\rho_A,\sigma_a\}\right)\right].
\end{equation*}
This relation also holds for identity measurement settings.

For brevity, let $X_a:=\Gamma_{BC|A}(\{\rho_A,\sigma_a\})$. Using the two-qubit Pauli completeness relation $\sum_{b,c=0}^{3}\operatorname{Tr}[(\sigma_b\otimes\sigma_c)X]\,\sigma_b\otimes\sigma_c=4X$, we find $\sum_{b,c=0}^{3}r_{abc}\sigma_b\otimes\sigma_c=2X_a$. Substitution into the Pauli expansion \eqref{R_ABC} then yields
\begin{align*}
R_{ABC}
&=\frac18\sum_{a=0}^{3}\sigma_a\otimes\left(\sum_{b,c=0}^{3}r_{abc}\sigma_b\otimes\sigma_c\right)\\
&=\frac14\sum_{a=0}^{3}\sigma_a\otimes X_a,
\end{align*}
which is precisely Eq.~\eqref{eq:pseudo_fork_PDO}.
\end{proof}

\section{Proof of Theorem~\ref{thm:pseudo_fork}}
\label{app:fork_proof}
\begin{proof}
Suppose first that $R_{ABC}$ admits a pseudo-Markov fork representation with common parent $A$. Let $(T_B,\boldsymbol p)$ and $(T_C,\boldsymbol q)$ be the Bloch-affine parameters of the two CPTP pseudo-channels $\mathcal E_{B|A}$ and $\mathcal E_{C|A}$, respectively. Their action on the Pauli basis is specified by
\begin{align*}
\mathcal E_{B|A}(I)
&=I+\boldsymbol p\cdot\boldsymbol\sigma,
&\mathcal E_{B|A}(\sigma_a)
&=\sum_b(T_B)_{ba}\sigma_b,\\
\mathcal E_{C|A}(I)
&=I+\boldsymbol q\cdot\boldsymbol\sigma,
&\mathcal E_{C|A}(\sigma_a)
&=\sum_c(T_C)_{ca}\sigma_c.
\end{align*}
All coefficients are real because the maps are Hermiticity preserving.

We begin with the parent--child marginals. By Lemma~\ref{lem:pseudo-channel-reconstruction}, compatibility with $R_{AB}$ and $R_{AC}$ gives
\begin{align*}
M&=T_B^{\mathsf T}
+\boldsymbol u\boldsymbol p^{\mathsf T},
&\boldsymbol v&=T_B\boldsymbol u+\boldsymbol p,\\
K&=T_C^{\mathsf T}+\boldsymbol u\boldsymbol q^{\mathsf T},
&\boldsymbol w&=T_C\boldsymbol u+\boldsymbol q.
\end{align*}
The correlation matrices therefore determine the linear parts of the two maps in terms of their translation vectors:
\begin{equation*}
T_B=M^{\mathsf T}-\boldsymbol p\boldsymbol u^{\mathsf T},~~T_C=K^{\mathsf T}-\boldsymbol q\boldsymbol u^{\mathsf T}.
\end{equation*}
Substituting these expressions into the relations for $\boldsymbol v$ and $\boldsymbol w$ yields
\begin{equation*}
\begin{aligned}
\boldsymbol v-M^{\mathsf T}\boldsymbol u
&=d_A\boldsymbol p,\\
\boldsymbol w-K^{\mathsf T}\boldsymbol u
&=d_A\boldsymbol q,
\end{aligned}
\end{equation*}
where $d_A=1-\|\boldsymbol u\|^2$. This proves Eq.~\eqref{eq:fork_F0}.

We next impose the common-parent factorization. In the unnormalized Choi convention, the conditional operators are
\begin{align*}
J_{B|A}
&=\frac12\left[I\otimes(I+\boldsymbol p\cdot\boldsymbol\sigma)+\sum_{a,b}(T_B)_{ba}\sigma_a^{\mathsf T}\otimes\sigma_b\right],\\
J_{C|A}
&=\frac12\left[I\otimes(I+\boldsymbol q\cdot\boldsymbol\sigma)+\sum_{a,c}(T_C)_{ca}\sigma_a^{\mathsf T}\otimes\sigma_c\right].
\end{align*}
As in the main text, these operators are understood to be embedded in the common $A^{\mathrm{out}}B^{\mathrm{in}}C^{\mathrm{in}}$ space, with spectator identities suppressed.

Only the terms acting nontrivially on the shared parent space can contribute to their commutator. Using $[\sigma_a^{\mathsf T},\sigma_d^{\mathsf T}] =-2i\sum_e\epsilon_{ade}\sigma_e^{\mathsf T}$, we find
\begin{equation*}
[J_{B|A},J_{C|A}]=-\frac{i}{2}\sum_{b,c,e}(\boldsymbol a_b\times\boldsymbol b_c)_e\,\sigma_e^{\mathsf T}\otimes\sigma_b\otimes\sigma_c,
\end{equation*}
where $\boldsymbol a_b=(T_B)_{b:}^{\mathsf T}$ and $\boldsymbol b_c=(T_C)_{c:}^{\mathsf T}$ are the rows of the Bloch matrices regarded as vectors in $\mathbb R^3$. Since the Pauli tensor products are linearly independent, the required commutativity $[J_{B|A},J_{C|A}]=0$ is equivalent to $\boldsymbol a_b\times\boldsymbol b_c=0$ for every $b,c$.

The marginal reconstruction relations give $\boldsymbol a_b=M_{:b}-p_b\boldsymbol u$ and $\boldsymbol b_c=K_{:c}-q_c\boldsymbol u$. Consequently,
\begin{equation*}
\begin{aligned}
0={}&(M_{:b}-p_b\boldsymbol u)\times(K_{:c}-q_c\boldsymbol u)\\
={}&M_{:b}\times K_{:c}-q_cM_{:b}\times\boldsymbol u-p_b\boldsymbol u\times K_{:c},
\end{aligned}
\end{equation*}
where the term proportional to $p_bq_c$ vanishes because $\boldsymbol u\times\boldsymbol u=0$. This establishes Eq.~\eqref{eq:fork_F1}.

To obtain the remaining conditions, we determine the joint pseudo-channel associated with $J_{BC|A}=J_{B|A}J_{C|A}$. The product of transposed Pauli operators obeys $\sigma_a^{\mathsf T}\sigma_d^{\mathsf T}=\delta_{ad}I-i\sum_e\epsilon_{ade}\sigma_e^{\mathsf T}$. The antisymmetric contribution vanishes by the commutativity condition derived above. Comparing the resulting product with the Choi expansion $J_{BC|A}=\frac12\sum_{\mu=0}^{3}\sigma_\mu^{\mathsf T}\otimes\Gamma_{BC|A}(\sigma_\mu)$,
we obtain
\begin{align}
\Gamma_{BC|A}(I)
={}&\frac12\Big[I\otimes I+\sum_b p_b\sigma_b\otimes I+\sum_c q_c I\otimes\sigma_c\nonumber\\
&\hspace{15mm}+\sum_{b,c}S_{bc}\,\sigma_b\otimes\sigma_c\Big],
\label{eq:app_fork_GammaI}\\
\Gamma_{BC|A}(\sigma_a)
={}&\frac12\Big[\sum_b(T_B)_{ba}\sigma_b\otimes I+\sum_c(T_C)_{ca}I\otimes\sigma_c\nonumber\\
&\hspace{15mm}+\sum_{b,c}(L_a)_{bc}\,\sigma_b\otimes\sigma_c\Big],
\label{eq:app_fork_GammaSigma}
\end{align}
where $S=\boldsymbol p\boldsymbol q^{\mathsf T} +T_BT_C^{\mathsf T}$ and $L_a=(T_B)_{:a}\boldsymbol q^{\mathsf T} +\boldsymbol p(T_C)_{:a}^{\mathsf T}$.

These expressions also verify the trace-preserving properties of the joint map: $\operatorname{Tr}\Gamma_{BC|A}(I)=2$ and $\operatorname{Tr}\Gamma_{BC|A}(\sigma_a)=0$. Moreover, tracing over either output system recovers the corresponding local pseudo-channel, namely, $\operatorname{Tr}_C\Gamma_{BC|A}=\mathcal E_{B|A}$ and $\operatorname{Tr}_B\Gamma_{BC|A}=\mathcal E_{C|A}$. Thus $\Gamma_{BC|A}$ is an HPTP joint pseudo-channel with the required marginals and common-parent factorization.

We now examine the child--child correlations. Tracing out the parent event in Eq.~\eqref{eq:pseudo_fork_PDO} gives $\rho_{BC}=\Gamma_{BC|A}(\rho_A)$. Using $\rho_A=(I+\boldsymbol u\cdot\boldsymbol\sigma)/2$ and Eqs.~\eqref{eq:app_fork_GammaI} and \eqref{eq:app_fork_GammaSigma}, the correlation matrix of $\rho_{BC}$ is
\begin{equation*}
N=S+(T_B\boldsymbol u)\boldsymbol q^{\mathsf T}+\boldsymbol p(T_C\boldsymbol u)^{\mathsf T}.
\end{equation*}
On the other hand, the two parent--child reconstruction relations imply
\begin{equation*}
\begin{aligned}
M^{\mathsf T}K={}
&(T_B+\boldsymbol p\boldsymbol u^{\mathsf T})(T_C^{\mathsf T}+\boldsymbol u\boldsymbol q^{\mathsf T})\\
={}&T_BT_C^{\mathsf T}+(T_B\boldsymbol u)\boldsymbol q^{\mathsf T}\\
&+\boldsymbol p(T_C\boldsymbol u)^{\mathsf T}+\|\boldsymbol u\|^2\boldsymbol p\boldsymbol q^{\mathsf T}.
\end{aligned}
\end{equation*}
Since $S=T_BT_C^{\mathsf T}+\boldsymbol p\boldsymbol q^{\mathsf T}$, subtracting these expressions yields
\begin{equation}
N-M^{\mathsf T}K=d_A\boldsymbol p\boldsymbol q^{\mathsf T}.
\label{eq:app_fork_N3}
\end{equation}
By Eq.~\eqref{eq:fork_F0}, $d_A\boldsymbol p=\boldsymbol v-M^{\mathsf T}\boldsymbol u$ and $d_A\boldsymbol q=\boldsymbol w-K^{\mathsf T}\boldsymbol u$. Equation~\eqref{eq:app_fork_N3} therefore becomes
\begin{equation*}
\begin{aligned}
N-M^{\mathsf T}K
&=(\boldsymbol v-M^{\mathsf T}\boldsymbol u)\boldsymbol q^{\mathsf T}\\
&=\boldsymbol p(\boldsymbol w-K^{\mathsf T}\boldsymbol u)^{\mathsf T},
\end{aligned}
\end{equation*}
which proves Eq.~\eqref{eq:fork_F2}.

It remains to consider the genuine three-body correlations. For $a=1,2,3$, the identity $\{\rho_A,\sigma_a\}=\sigma_a+u_aI$ and the fork PDO representation give
\begin{equation*}
\begin{aligned}
G_{abc}
&=\frac12\operatorname{Tr}\left[(\sigma_b\otimes\sigma_c)\Gamma_{BC|A}(\{\rho_A,\sigma_a\})\right]\\
&=u_aS_{bc}+(L_a)_{bc}.
\end{aligned}
\end{equation*}
Writing $G_a=[G_{abc}]_{b,c=1}^{3}$, we have $G_a=u_aS+(T_B)_{:a}\boldsymbol q^{\mathsf T}+\boldsymbol p(T_C)_{:a}^{\mathsf T}$.

To eliminate the remaining channel parameters, we use $T_B\boldsymbol u=\boldsymbol v-\boldsymbol p$ and $T_C\boldsymbol u=\boldsymbol w-\boldsymbol q$. The expression for $N$ then gives
\begin{equation*}
S=N-\boldsymbol v\boldsymbol q^{\mathsf T}-\boldsymbol p\boldsymbol w^{\mathsf T}+2\boldsymbol p\boldsymbol q^{\mathsf T}.
\end{equation*}
Furthermore, $(T_B)_{:a}=M_{a:}^{\mathsf T} -u_a\boldsymbol p$
and $(T_C)_{:a}=K_{a:}^{\mathsf T}-u_a\boldsymbol q$. Substituting these relations into the expression for $G_a$, the terms proportional to $\boldsymbol p\boldsymbol q^{\mathsf T}$ cancel, leaving
\begin{equation*}
\begin{aligned}
G_a={}&u_aN+(M_{a:}^{\mathsf T}-u_a\boldsymbol v)\boldsymbol q^{\mathsf T}\\
&+\boldsymbol p(K_{a:}-u_a\boldsymbol w^{\mathsf T}).
\end{aligned}
\end{equation*}
In components, this is $G_{abc}=u_aN_{bc}+(M_{ab}-u_av_b)q_c+p_b(K_{ac}-u_aw_c)$, which is precisely Eq.~\eqref{eq:fork_F3}. Necessity is established.

Conversely, suppose that there exist $\boldsymbol p,\boldsymbol q\in\mathbb R^3$ satisfying Eqs.~\eqref{eq:fork_F0}--\eqref{eq:fork_F3}. We construct the two outgoing pseudo-channels and show that their common-cause extension reproduces the prescribed PDO.

Define $T_B=M^{\mathsf T}-\boldsymbol p\boldsymbol u^{\mathsf T}$ and $T_C=K^{\mathsf T}-\boldsymbol q\boldsymbol u^{\mathsf T}$, and use these matrices and translation vectors to specify the action of $\mathcal E_{B|A}$ and $\mathcal E_{C|A}$ on the Pauli basis, as at the beginning of the proof. Since the coefficients are real, both maps preserve Hermiticity. They also preserve trace, because their images of $I$ have trace $2$, whereas their images of $\sigma_a$ are traceless. Thus both maps are HPTP pseudo-channels.

By construction, $M=T_B^{\mathsf T}+\boldsymbol u\boldsymbol p^{\mathsf T}$ and $K=T_C^{\mathsf T}+\boldsymbol u\boldsymbol q^{\mathsf T}$. Furthermore, Eq.~\eqref{eq:fork_F0} gives
\begin{equation*}
\begin{aligned}
T_B\boldsymbol u+\boldsymbol p
&=M^{\mathsf T}\boldsymbol u+d_A\boldsymbol p=\boldsymbol v,\\
T_C\boldsymbol u+\boldsymbol q
&=K^{\mathsf T}\boldsymbol u+d_A\boldsymbol q=\boldsymbol w.
\end{aligned}
\end{equation*} 
Hence, the constructed maps reproduce the prescribed $AB$ and $AC$ marginals.

We next verify their common-cause compatibility. By Eq.~\eqref{eq:fork_F1}, the vectors $\boldsymbol a_b=M_{:b}-p_b\boldsymbol u$ and $\boldsymbol b_c=K_{:c}-q_c\boldsymbol u$ satisfy $\boldsymbol a_b\times\boldsymbol b_c=0$ for every $b,c$. The commutator expression derived above therefore implies $[J_{B|A},J_{C|A}]=0$. Consequently, the product $J_{BC|A}=J_{B|A}J_{C|A}$ is Hermitian and determines a joint HPTP pseudo-channel $\Gamma_{BC|A}$ through Eqs.~\eqref{eq:app_fork_GammaI} and \eqref{eq:app_fork_GammaSigma}.

Let $\widetilde R_{ABC}$ be the PDO generated by $\rho_A$ and this joint pseudo-channel through Eq.~\eqref{eq:pseudo_fork_PDO}. Its parent--child marginals already agree with those of $R_{ABC}$. For the child--child sector, Eq.~\eqref{eq:app_fork_N3} gives
\begin{equation*}
\widetilde N=M^{\mathsf T}K+d_A\boldsymbol p\boldsymbol q^{\mathsf T}.
\end{equation*}
Using Eq.~\eqref{eq:fork_F0}, condition \eqref{eq:fork_F2} ensures that $\widetilde N=N$.

Finally, the three-body coefficients of the constructed PDO obey the relation derived in the necessity part,
\begin{equation*}
\begin{aligned}
\widetilde G_{abc}={}&u_a\widetilde N_{bc}+(M_{ab}-u_av_b)q_c\\
&+p_b(K_{ac}-u_aw_c).
\end{aligned}
\end{equation*}
Since $\widetilde N=N$, Eq.~\eqref{eq:fork_F3} implies $\widetilde G_{abc}=G_{abc}$ for all $a,b,c=1,2,3$.

The constructed PDO therefore has exactly the same one-body, two-body, and three-body Pauli coefficients as $R_{ABC}$. By completeness of the three-qubit Pauli basis, $\widetilde R_{ABC}=R_{ABC}$. This establishes the existence of a pseudo-Markov fork representation and completes the proof.
\end{proof}

\section{Proof of Proposition~\ref{prop:collider_PDO}}
\label{app:collider_representation}
\begin{proof}
For the collider structure $A\to C\leftarrow B$, the two root systems are independently prepared in the product state $\rho_A\otimes\rho_B$. Applying Eq.~\eqref{eq:measurement_identity} at both root nodes, we obtain
\begin{equation*}
\begin{aligned}
&\sum_{x,y=\pm1}xy\,\Pi_{x|a}\rho_A\Pi_{x|a}\otimes\Pi_{y|b}\rho_B\Pi_{y|b}\\
&\qquad=\frac14\{\rho_A,\sigma_a\}\otimes\{\rho_B,\sigma_b\}.
\end{aligned}
\end{equation*}
This identity also holds when either measurement setting is the identity setting. Using $\sum_{z=\pm1}z\Pi_{z|c}=\sigma_c$ for the measurement at $C$, the corresponding Pauli coefficients are
\begin{equation*}
r_{abc}=\frac14\operatorname{Tr}\left[\sigma_c\,\mathcal E_{C|AB}\left(\{\rho_A,\sigma_a\}\otimes\{\rho_B,\sigma_b\}\right)\right].
\end{equation*}
The Pauli completeness relation $\sum_{c=0}^{3}\operatorname{Tr}(\sigma_c X)\sigma_c=2X$ then gives
\begin{equation*}
\begin{aligned}
R_{ABC}
&=\frac18\sum_{a,b=0}^{3}\sigma_a\otimes\sigma_b\otimes\left(\sum_{c=0}^{3}r_{abc}\sigma_c\right)\\
&=\frac1{16}\sum_{a,b=0}^{3}\sigma_a\otimes\sigma_b\otimes\mathcal E_{C|AB}\left(\{\rho_A,\sigma_a\}\otimes\{\rho_B,\sigma_b\}\right),
\end{aligned}
\end{equation*}
which establishes Eq.~\eqref{eq:collider_PDO}.
\end{proof}

\section{Proof of Theorem~\ref{thm:collider}}
\label{app:collider_proof}
\begin{proof}
We first establish necessity. Suppose that $R_{ABC}$ admits a pseudo-Markov collider representation $A\to C\leftarrow B$, with independently prepared root states $\rho_A$ and $\rho_B$ and an HPTP joint pseudo-channel $\mathcal E_{C|AB}$.

For a qubit state with Bloch vector $\boldsymbol r$, the Pauli anticommutation relations give $\{\rho,\sigma_a\}=\sum_{\mu=0}^{3}H(\boldsymbol r)_{\mu a}\sigma_\mu$. Combining this identity with the collider PDO representation \eqref{eq:collider_PDO} and the Pauli-transfer expansion \eqref{eq:collider_Pauli_transfer}, we obtain
\begin{equation*}
r_{abc}=\sum_{\mu,\nu=0}^{3}H(\boldsymbol u)_{\mu a}\ell_{c|\mu\nu}H(\boldsymbol v)_{\nu b}.
\end{equation*}
In matrix form,
\begin{equation*}
\mathcal R^{(c)}=H(\boldsymbol u)^{\mathsf T}L^{(c)}H(\boldsymbol v),~~c=0,1,2,3.
\end{equation*}

Trace preservation of $\mathcal E_{C|AB}$ requires $\ell_{0|\mu\nu}=\delta_{\mu0}\delta_{\nu0}$, so that $L^{(0)}=E_{00}$. Consequently,
\begin{equation*}
\begin{aligned}
\mathcal R^{(0)}
&=H(\boldsymbol u)^{\mathsf T}E_{00}H(\boldsymbol v)\\
&=
\begin{pmatrix}
1\\ \boldsymbol u
\end{pmatrix}
\begin{pmatrix}
1 & \boldsymbol v^{\mathsf T}
\end{pmatrix}.
\end{aligned}
\end{equation*}
Comparison with Eq.~\eqref{eq:collider_R0} yields $M=\boldsymbol u\boldsymbol v^{\mathsf T}$, which is Eq.~\eqref{eq:collider_root_independence}. For $c=1,2,3$, the remaining matrix relations give Eq.~\eqref{eq:collider_feasibility}. This establishes necessity.

Conversely, suppose that Eq.~\eqref{eq:collider_root_independence} holds and that real matrices $L^{(1)},L^{(2)},L^{(3)}$ satisfy Eq.~\eqref{eq:collider_feasibility}. Set $L^{(0)}=E_{00}$ and define a linear map $\mathcal E_{C|AB}$ on the two-qubit Pauli basis by
\begin{equation*}
\mathcal E_{C|AB}(\sigma_\mu\otimes\sigma_\nu)=2\sum_{c=0}^{3}\ell_{c|\mu\nu}\sigma_c.
\end{equation*}
Since all coefficients are real, the map preserves Hermiticity. Furthermore,
\begin{equation*}
\begin{aligned}
\operatorname{Tr}\mathcal E_{C|AB}(\sigma_\mu\otimes\sigma_\nu)
&=4\ell_{0|\mu\nu}\\
&=4\delta_{\mu0}\delta_{\nu0}\\
&=\operatorname{Tr}(\sigma_\mu\otimes\sigma_\nu).
\end{aligned}
\end{equation*}
As the Pauli products form an operator basis, this establishes trace preservation for arbitrary inputs. Hence $\mathcal E_{C|AB}$ is HPTP.

Let $\widetilde R_{ABC}$ be the PDO generated by the product root state $\rho_A\otimes\rho_B$ and the constructed joint pseudo-channel through Eq.~\eqref{eq:collider_PDO}. Its Pauli coefficients satisfy
\begin{equation*}
\widetilde{\mathcal R}^{(c)}=H(\boldsymbol u)^{\mathsf T}L^{(c)}H(\boldsymbol v),~~ c=0,1,2,3.
\end{equation*}
For $c=0$, the choice $L^{(0)}=E_{00}$ and Eq.~\eqref{eq:collider_root_independence} ensure that $\widetilde{\mathcal R}^{(0)} =\mathcal R^{(0)}$. For $c=1,2,3$, the same equality follows from Eq.~\eqref{eq:collider_feasibility}.

Thus, $\widetilde R_{ABC}$ reproduces all Pauli coefficients of $R_{ABC}$. By completeness of the three-qubit Pauli basis, $\widetilde R_{ABC}=R_{ABC}$. The constructed HPTP map therefore provides the required pseudo-Markov collider representation, proving sufficiency.
\end{proof}

\section{Proof of Corollary~\ref{cor:collider_reconstruction}}
\label{app:collider_reconstruction_proof}
\begin{proof}
For each $c=1,2,3$, Theorem~\ref{thm:collider} reduces the reconstruction problem to the matrix equation
\begin{equation*}
    H(\boldsymbol u)L^{(c)}H(\boldsymbol v) =\mathcal R^{(c)}.
\end{equation*}
We use the standard solution of the matrix equation $AXB=C$ in terms of Moore--Penrose inverses: whenever the equation is consistent, its complete solution set is
\begin{equation*}
    X=A^{+}CB^{+}+Y -A^{+}AYBB^{+},
\end{equation*}
where $Y$ is arbitrary.

Applying this result to $A=H(\boldsymbol u)$, $B=H(\boldsymbol v)$, and $C=\mathcal R^{(c)}$ immediately gives
\begin{align*}
    L^{(c)}={}&H(\boldsymbol u)^{+}\mathcal R^{(c)}H(\boldsymbol v)^{+}+Y_c\nonumber\\
    &-H(\boldsymbol u)^{+}H(\boldsymbol u)Y_cH(\boldsymbol v)H(\boldsymbol v)^{+}.
\end{align*}

\end{proof}

\bibliography{apssamp}

\end{document}